\documentclass[preprint, 12pt]{elsarticle}
\usepackage{hyperref}
\usepackage{lineno}
\usepackage{amsmath}
\usepackage{subfigure}

\usepackage{color}
\usepackage{longtable}
\usepackage{amsmath,amssymb}

\usepackage{threeparttable}

\modulolinenumbers[1]
\usepackage{docmute}

\newcommand*\patchAmsMathEnvironmentForLineno[1]{
  \expandafter\let\csname old#1\expandafter\endcsname\csname #1\endcsname
  \expandafter\let\csname oldend#1\expandafter\endcsname\csname end#1\endcsname
  \renewenvironment{#1}
     {\linenomath\csname old#1\endcsname}
     {\csname oldend#1\endcsname\endlinenomath}}
\newcommand*\patchBothAmsMathEnvironmentsForLineno[1]{
  \patchAmsMathEnvironmentForLineno{#1}
  \patchAmsMathEnvironmentForLineno{#1*}}
\AtBeginDocument{
\patchBothAmsMathEnvironmentsForLineno{equation}
\patchBothAmsMathEnvironmentsForLineno{align}
\patchBothAmsMathEnvironmentsForLineno{flalign}
\patchBothAmsMathEnvironmentsForLineno{alignat}
\patchBothAmsMathEnvironmentsForLineno{gather}
\patchBothAmsMathEnvironmentsForLineno{multline}
}

\usepackage{setspace}

\begin{document}
\begin{frontmatter}

\title{Development of the X-ray polarimeter using CMOS imager: polarization sensitivity of a $1.5~{\rm \mu m}$ pixel CMOS sensor}

\author[utokyo]{Toshiya Iwata\corref{cor1}}
\author[utokyo]{Kouichi Hagino}
\author[utokyo,osakau,ipmu]{Hirokazu Odaka}
\author[isas,utokyo]{Tsubasa Tamba}
\author[utokyo]{Masahiro Ichihashi}
\author[utokyo]{Tatsuaki Kato}
\author[osakau]{Kota Ishiwata}
\author[osakau]{Haruki Kuramoto}
\author[utokyo]{Hiroumi Matsuhashi}
\author[utokyo]{Shota Arai}
\author[utokyo]{Takahiro Minami}
\author[utokyo]{Satoshi Takashima}
\author[utokyo,resceu,tsqsi]{Aya Bamba}

\address[utokyo]{Department of Physics, Faculty of Science, The University of Tokyo, 7-3-1 Hongo, Bunkyo-ku, Tokyo 113-0033, Japan}
\address[isas]{Japan Aerospace Exploration Agency, Institute of Space and Astronautical Science, 3-1-1 Yoshino-dai, Chuo-ku, Sagamihara, Kanagawa 252-5210, Japan}
\address[osakau]{Department of Earth and Space Science, Osaka University, 1-1 Machikaneyama-cho, Toyonaka, Osaka 560-0043, Japan}
\address[ipmu]{Kavli IPMU, The University of Tokyo, Kashiwa 113-0033, Japan}
\address[resceu]{Research Center for Early Universe, Faculty of Science, The University of Tokyo, 7-3-1 Hongo, Bunkyo-ku, Tokyo 113-0033, Japan}
\address[tsqsi]{Trans-Scale Quantum Science Institute, The University of Tokyo, Tokyo  113-0033, Japan}

\cortext[cor1]{Email Address: toshiya.iwata@phys.s.u-tokyo.ac.jp}

\begin{abstract}
We are developing an imaging polarimeter by combining a fine-pixel CMOS image sensor with a coded aperture mask as part of the cipher project, aiming to achieve X-ray polarimetry in the energy range of $10$--$30~\mathrm{keV}$. A successful proof-of-concept experiment was conducted using a fine-pixel CMOS sensor with a $2.5~\mathrm{\mu m}$ pixel size. In this study, we conducted beam experiments to assess the modulation factor (MF) of the CMOS sensor with a $1.5~\mathrm{\mu m}$ pixel size manufactured by Canon and to determine if there was any improvement in the MF. The measured MF was $8.32\% \pm 0.34\%$ at $10~\mathrm{keV}$ and $16.10\% \pm 0.68\%$ at $22~\mathrm{keV}$, exceeding those of the $2.5~\mathrm{\mu m}$ sensor in the $6$--$22~\mathrm{keV}$ range. We also evaluated the quantum efficiency of the sensor, inferring a detection layer thickness of $2.67 \pm 0.48~{\rm \mu m}$. To develop a more sensitive polarimeter, a sensor with a thicker detection layer, smaller pixel size, and reduced thermal diffusion effect is desirable.
\end{abstract}

\end{frontmatter}


\section{Introduction}
\label{sec_intro}

X-ray polarimetry of astronomical objects holds promise for understanding magnetic field structures, the geometry of scattering materials, and the curvature of spacetime around black holes. Instruments like PoGO+ \cite{chauvin2017} and the Hitomi Soft Gamma-ray Detector \cite{hitomi2018} have conducted observations of polarization in hard X-rays and soft gamma rays above $20~\mathrm{keV}$. The Imaging X-ray Polarimetry Explorer (IXPE; \cite{weisskopf2022}), launched in December 2021, is performing polarimetry in the soft X-ray range ($2$--$8~\mathrm{keV}$). The energy range between these, $10$--$30~\mathrm{keV}$, is crucial for X-ray polarimetry due to the dominance of non-thermal radiation and the abundance of photons. However, X-ray polarimetry in the energy range of $10$--$30~\mathrm{keV}$ remains unexplored. X-Calibur addresses this gap by employing a low-Z scattering element for its Compton polarimeter \cite{beilicke2014, abarr2020}.

We presented an alternative approach utilizing photoionization in silicon, named cipher \cite{odaka2020}. By employing silicon detectors, we anticipate an enhancement in the quantum efficiency of hard X-ray due to approximately $10^3$ times higher density of detector material compared to gas pixel detectors used in IXPE. Moreover, improved energy resolution, with a lower average energy required to produce an electron-hole pair ($\simeq 3.6~\mathrm{eV}$), will enable simultaneous X-ray polarimetry and spectroscopy. Our aim is to achieve imaging polarimetry by combining a fine-pixel CMOS image sensor with a coded aperture mask. In the dominant K-shell photoionization process, the cross-section is proportional to $\cos^2{\phi}$, where $\phi$ is the azimuthal angle of the emitted photoelectron relative to the polarization direction of the incident X-ray photons. By analyzing the charge distribution of events detected by the fine-pixel sensor, we can estimate the photoelectron emission angle. Consequently, we can deduce the polarization degree and polarization angle of the incident X-rays. In a proof-of-concept experiment using a fine-pixel CMOS sensor with a pixel size of $2.5~\mathrm{\mu m}$ \cite{odaka2020}, we successfully measured the modulation factor (MF), which is the observed modulation amplitude for a fully polarized incident beam. Additionally, we have achieved imaging using a coded aperture mask that reduces imaging artifacts by combining multiple random patterns \cite{kasuga2020}, and further improved imaging polarimetry by employing a reconstruction method with the expectation-maximization algorithm (Tamba et al. submitted). However, enhancing the MF of the detector remains a critical objective for improving polarization sensitivity.

The MF of a detector is influenced by several sensor properties, including pixel size and thermal diffusion effects. A smaller pixel size sensor allows for more precise tracking of photoelectrons, potentially enhancing the MF. However, significant diffusion effects can erase information on the initial charge distribution, leading to a decrease in the MF. In this study, we conducted beam experiments utilizing a CMOS sensor with a pixel size of $1.5~\mathrm{\mu m}$ manufactured by Canon to investigate if the MF is improved. Additionally, we evaluate the quantum efficiency (QE) of the sensor to assess its polarization sensitivity. Section \ref{sec_beam} describes the setup and procedure of the beam experiments. In section \ref{sec_results}, we outline the analysis method used to determine the MF and QE of the sensor, comparing the obtained results with those of the $2.5~\mathrm{\mu m}$ sensor. We delve into the discussion of polarization sensitivity in section \ref{sec_discussion} and summarize our conclusions in section \ref{sec_conclusion}.

\section{X-ray beam experiment}
\label{sec_beam}

To assess the MF and QE of the CMOS sensors, we performed beam experiments at the SPring-8 synchrotron radiation facility and the Photon Factory of High Energy Accelerator Research Organization (KEK-PF) in Japan.
In subsection \ref{sub_sensors}, we detail the sensors utilized in the beam experiments. Subsections \ref{sub_sp8} and \ref{sub_kek} outline the setup and procedures employed during the beam experiments at SPring-8 and KEK-PF, respectively.

\subsection{Fine-pixel CMOS image sensor LI8020SA}
\label{sub_sensors}

The LI8020SA is a front-illuminated CMOS image sensor manufactured by Canon, designed primarily for visible light imaging. The imaging area, spanning $29.35~\mathrm{mm} \times 18.88~\mathrm{mm}$, consists of $19568 \times 12588$ pixels, with each pixel measuring $1.5~{\rm \mu m} \times 1.5~{\rm \mu m}$. The properties of the LI8020SA sensor are detailed in Tab. \ref{tb:sensor_properties}.
In order to increase sensitivity to X-rays, the cover glass of the sensor was removed for experiments conducted at KEK-PF.

\begin{table}[ht!]
\caption{Properties of the CMOS sensor LI8020SA.}
\begin{center}
\begin{tabular}{ c c }
\hline \hline
 & LI8020SA  \\
\hline
Pixel size & $1.5~\mathrm{\mu m} \times 1.5~\mathrm{\mu m} $  \\
Number of pixels & $19568 \times 12588 $ \\
Imaging area & $29.35~\mathrm{mm} \times 18.88~\mathrm{mm}$ \\
\hline 
\end{tabular}
\label{tb:sensor_properties}
\end{center}
\end{table}

\subsection{SPring-8 BL20B2}
\label{sub_sp8}

To assess the MFs of the LI8020SA sensors, we conducted X-ray beam experiments at BL20B2 in SPring-8 \citep{goto2001} in June 2023. The photon beam of the beamline was monochromatic and nearly 100\% linearly polarized, with a reported polarization degree of $99.31\% \pm 0.03\%$ at $12.4~\mathrm{keV}$ \cite{asakura2019}.

\begin{figure}[htb]
\begin{center}
\includegraphics[width=8cm]{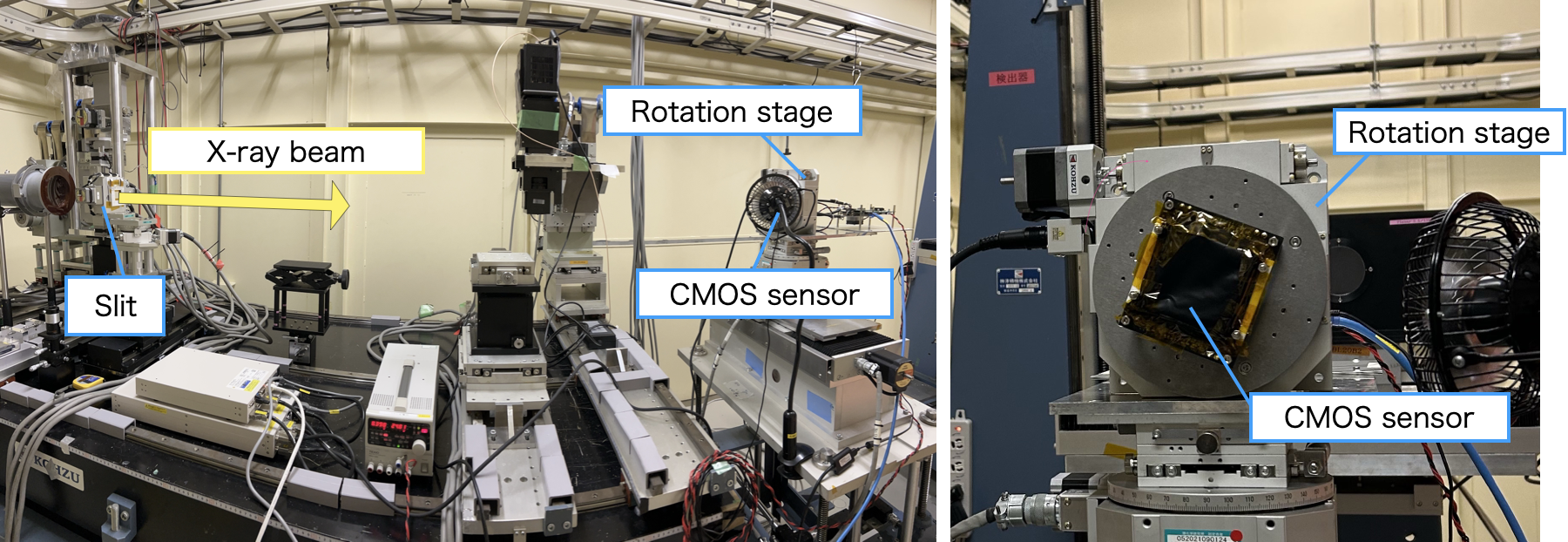}
\caption{Experimental setup at BL20B2 in SPring-8 (left) and the setup as seen from the upstream of the beam (right). The sensor is attached to the rotation stage and is covered with a dark curtain. Attenuators are inserted downstream of the slit.}
\label{fig:sp8_2023}
\end{center}
\end{figure}

During the SPring-8 beam experiment, the LI8020SA CMOS sensor was exposed to the X-ray beam using the setup depicted in Fig. \ref{fig:sp8_2023}. We conducted measurements at beam energies ranging from $10$ to $22~\mathrm{keV}$, with the beam size collimated by a slit to approximately $10~\mathrm{mm} \times 10~\mathrm{mm}$. Al and Cu plates were inserted as attenuators to prevent event pile-up. The sensor was mounted on a rotation stage, aligning the rotation axis with the center of the CMOS sensor. We covered the sensor with a dark curtain to block visible light and cooled it with an electric fan to keep the temperature of the sensor at about $30^{\circ}\mathrm{C}$--$40^{\circ}\mathrm{C}$. 

Frame data were acquired at various polarization angles from $0$, $45$, and $90$ degrees by rotating the stage. The exposure time for each frame was $1.613$~s per frame. Additionally, we acquired data without X-ray beam irradiation at each energy to identify noisy pixels. The data acquisition conditions are summarized in Tab. \ref{tb:beam_exp_conditions_sp8}.

\begin{table*}[htbp]
\centering
\caption{List of data acquired at SPring-8.
}
\begin{tabular}{ccccc}
\hline \hline
 Beam energy  & Polarization angle & Frame exposure & Number of frames & gain\\
\hline
 10.0~keV& $0^{\circ}$, $90^{\circ}$, $45^{\circ}$ & 1613~ms & 104, 104, 104 & 1 \\
 12.0~keV& $0^{\circ}$, $90^{\circ}$, $45^{\circ}$ & 1613~ms & 104, 104, 104 & 1\\
 16.0~keV& $0^{\circ}$, $90^{\circ}$, $45^{\circ}$ & 1613~ms & 104, 104, 104 & 1\\
 22.0~keV& $0^{\circ}$, $90^{\circ}$, $45^{\circ}$ & 1613~ms & 104, 104, 104 & 1/2\\
\hline
\end{tabular}
\label{tb:beam_exp_conditions_sp8}
\end{table*}

\subsection{KEK-PF BL14A}
\label{sub_kek}

\begin{figure}[htb]
\begin{center}
\includegraphics[width=7cm]{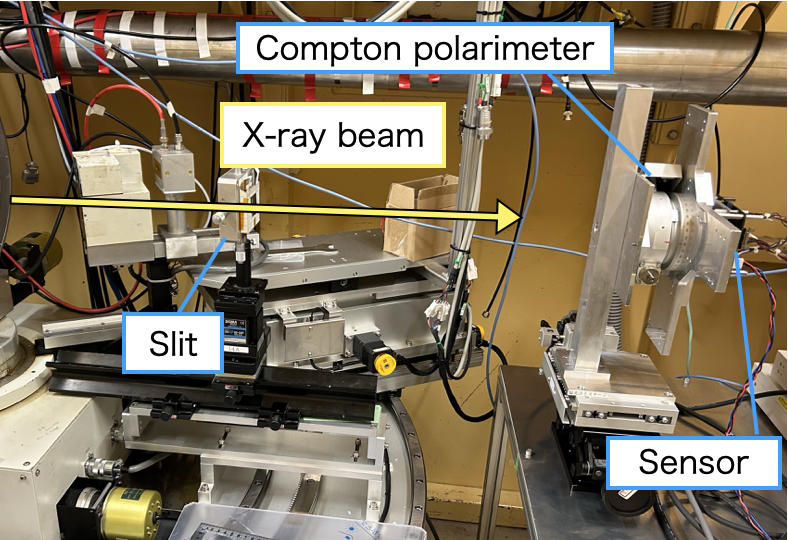}
\caption{Experimental setup at KEK-PF. The CMOS sensor is attached to the rotation stage in the Compton polarimeter system. A $2~\mathrm{mm}$ radius hole in the Compton polarimeter system acts as a collimator. Attenuators are inserted downstream of the slit.}
\label{fig:kek}
\end{center}
\end{figure}

To assess the MFs below $10~\mathrm{keV}$ and the QE of the LI8020SA, we conducted an X-ray beam experiment at BL-14A in KEK-PF in November 2023. The beam size was approximately $2~\mathrm{mm} \times 4~\mathrm{mm}$ and was collimated by a slit when necessary. Al and/or Cu attenuators were inserted downstream of the slit to control beam intensity. The transmitted beam underwent further collimation through a $2~\mathrm{mm}$ radius hole in the Compton polarimeter system, as illustrated in Figure \ref{fig:kek}.

Initially, we evaluated the polarization degree of the incident X-ray beam using a Compton polarimeter. The X-ray beam passed through a collimator hole and interacted with a beryllium scatterer placed at the end of the collimator. Photons scattered at $90$ degrees within the scatterer were detected by a silicon drift detector (SDD). By rotating the stage, we measured count rates at various azimuthal angles. The observed modulation of the count rate azimuthal distribution was $68.3\% \pm 0.4\%$ at $6~\mathrm{keV}$ and $69.8\% \pm 0.3\%$ at $10~\mathrm{keV}$. Dividing the modulation amplitude by the polarimeter's intrinsic MF yielded the polarization degree of the beam as $78.4\% \pm 0.9\%$ at $6~\mathrm{keV}$ and $78.9\% \pm 0.9\%$ at $10~\mathrm{keV}$. The intrinsic MF of the polarimeter was estimated through simulation using ComptonSoft based on Geant4 \cite{agostinelli2003, allison2006, allison2016}. 

Subsequently, we proceeded to measure the modulation amplitude of the LI8020SA sensor. We replaced the SDD and the scatterer with the LI8020SA sensor to allow irradiation by the incident beam. By rotating the stage, we collected frame data at polarization angles of $0$, $-45$, and $-90$ degrees. To block visible light, the sensor was shielded with aluminum foil, which had a thickness of $11~\mathrm{\mu m}$. The acquired data are summarized in Tab. \ref{tb:beam_exp_conditions_kek_mf}. The determination of the modulation amplitude and MF of the LI8020SA sensor is presented in section \ref{sec_results}.

To evaluate the QE of the LI8020SA sensor, we compared the count rates of the CMOS sensor with those of the SDD under identical conditions. The count rate measurements of the CMOS sensor were carried out using the same setup as that for the modulation amplitude measurements of the LI8020SA sensor. Afterward, we replaced the CMOS sensor with the SDD and exposed it to the same conditions as the CMOS sensor measurements. Both the CMOS sensor and the SDD were shielded with aluminum foil of $11~\mathrm{\mu m}$ thickness.  The acquired data are summarized in Tab. \ref{tb:beam_exp_conditions_kek_qe}.

\begin{table*}[htbp]
\centering
\caption{List of data acquired at KEK-PF for modulation factor measurement.
}
\begin{tabular}{ccccc}
\hline \hline
 Beam energy & Polarization angle & Frame exposure & Number of frames & gain\\
\hline
6.0~keV & $0^{\circ}$, $-90^{\circ}$, $-45^{\circ}$ & $645$~ms & 105, 105, 105 & 1\\
10.0~keV & $0^{\circ}$, $-90^{\circ}$, $-45^{\circ}$ & $645$~ms & 105, 105, 105 & 1\\
\hline
\end{tabular}
\label{tb:beam_exp_conditions_kek_mf}
\end{table*}

\begin{table*}[htbp]
\centering
\caption{List of data acquired at KEK-PF for quantum efficiency measurement.
}
\begin{threeparttable}
\begin{tabular}{cccc}
\hline \hline
Beam energy & gain  & Frame exposure  & Number of frames  \\
\hline
6.0~keV & 1     & 645~ms & 20         \\
10.0~keV & 1, 1/2${}^a$     & 645~ms & 20, 20${}^a$          \\
12.0~keV & 1, 1/2, 1/2, 1/2, 1/2${}^a$     & 645~ms & 30, 20, 20, 30, 30${}^a$  \\
16.0~keV & 1    & 645~ms & 85       \\
18.0~keV & 1/2    & 645~ms & 55      \\
\hline
\end{tabular}
\begin{tablenotes}
\small 
\item ${}^a$ We acquired multiple datasets to estimate the uncertainty in the quantum efficiency.
\end{tablenotes}
\end{threeparttable}
\label{tb:beam_exp_conditions_kek_qe}
\end{table*}

\section{Results}
\label{sec_results}

\subsection{X-ray event analysis procedure}
\label{sub_spectra}

X-ray events were extracted from the acquired data through the following procedure. Initially, we identified noisy pixels by analyzing the dark data, which refers to data collected without an X-ray beam. Specifically, we examined multiple frames and flagged pixels exhibiting excessively high mean and standard deviation in pixel values as noisy. 
Secondly, we perform dark-level subtraction on all frames. The dark levels are computed by averaging the pixel values of the four frames preceding the frame of interest. 

Subsequently, X-ray events are identified by scanning the entire frame for pixels that satisfy the following criteria: the signal values exceed an event threshold, the pixel has the maximum value within the $9 \times 9$ pixel event analysis area centered on the pixel, and the event analysis area does not contain any noisy pixels. 
We set the event threshold values high enough to reduce false event identification due to noise, yet low enough to avoid excluding X-ray events; specifically, it was set at $\simeq 0.9$~keV.

Finally, a photoelectron track image was defined for each identified event by examining the $9 \times 9$ pixels surrounding the event center. During this process, pixels with values exceeding a predefined split threshold were included in the track image. The split threshold value was chosen to optimize polarization sensitivity; specifically, it was set at $\simeq 0.09$~keV. We refer to the number of pixels within the track image as ``multiplicity", and the sum of all pixel values belonging to this event as ``reconstructed energy".

\subsection{Modulation factor}
\label{sub_mf}

The MF, a crucial parameter of a polarimeter, is defined as the modulation amplitude measured when a fully polarized beam interacts with the detector. It can be determined by fitting the distribution of photoelectron emission angles $\phi$ using the function
\begin{equation}
y = C\left[m \cos \left(2\phi - 2B\right) + 1 \right]
\label{eq:modulation_curve}
\end{equation}
where $m$ represents the modulation amplitude and $C$ and $B$ are constants. The MF is then calculated by dividing $m$ by the polarization degree of the incident beam.

To evaluate the MF of the sensor, we derived the modulation amplitude of the data through the following steps. Initially, we acquired the distribution of photoelectron emission angles at $0$, $45$, and $90$ degrees\footnote{We acquired data at $0^\circ$, $-45^\circ$, and $-90^\circ$ angles, which was convenient for the experimental setup at the KEK-PF experiment. Hence, we utilized these angle data for the KEK-PF dataset.}.
We estimated the photoelectron emission angle for each event based on the second moment of the pixel value distribution:
\begin{equation}
    M_2(\phi) = \frac{\sum_iQ_i\left(x_i\cos\phi + y_i\sin\phi\right)^2}{\sum_iQ_i}.
    \label{eq:m2}
\end{equation}
Here, $Q_i$ represents the pixel value of pixel $i$, and $x_i$ and $y_i$ represent the coordinates of the pixel $i$ with respect to the origin of the coordinate system, which is defined at the barycenter of the charge distribution \cite{bellazzini2003}. 
The angle providing the maximum value of $M_2$ was considered the photoelectron emission angle.
In this analysis, we used events within $\pm 2\sigma$ of the energy resolution around the reconstructed energy peak, corresponding to monochromatic X-ray beam energy.
Secondly, we canceled spurious modulation by normalizing the modulation curve by the sum of the $0$ degree and $90$ degree modulation curves. Subsequently, to mitigate geometrically anisotropic responses, we combined the normalized and angle-shifted $45$ degree modulation curve with the $0$ degree modulation curve, simulating a scenario with two sensors positioned at a $45$ degree rotation. Finally, we determined the MFs of the sensors by fitting the function \ref{eq:modulation_curve} to the obtained corrected modulation curve and dividing $m$ by the polarization degree of the incident beam.

\begin{figure}[htb]
\begin{center}
\includegraphics[width=7cm]{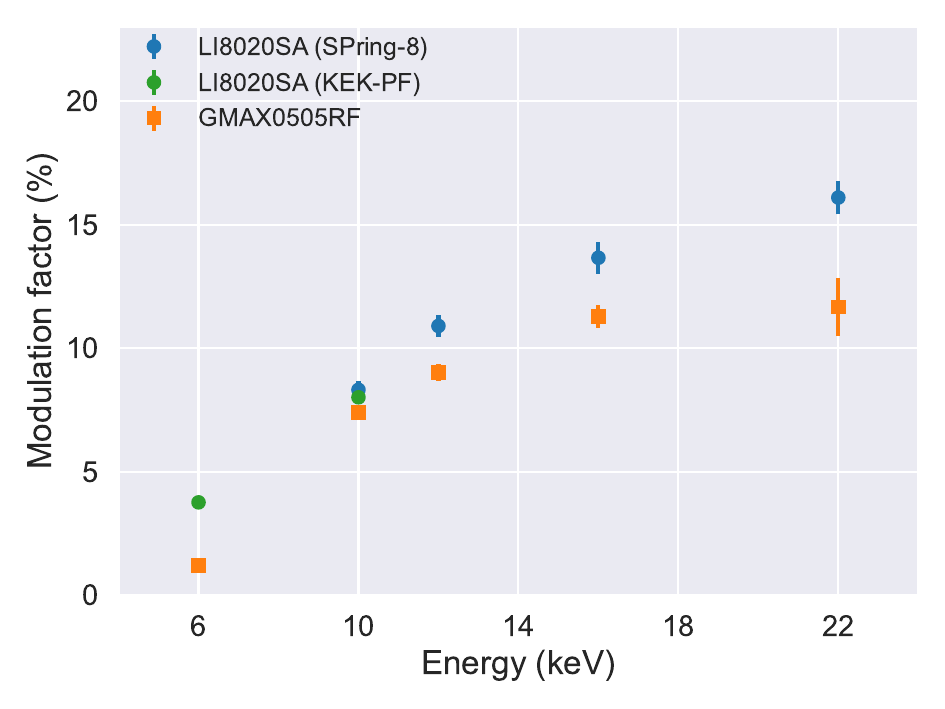}
\caption{Modulation factors of the CMOS sensors at each energy. The modulation factors of the GMAX0505RF were obtained from Odaka et al. (in prep). 
The events used for the analysis were within the multiplicity range of $2$--$6$ for LI8020SA and $2$--$4$ for GMAX0505RF.}
\label{fig:mf}
\end{center}
\end{figure}

The determined MFs are summarized in Tab. \ref{tb:mf} and illustrated in Fig. \ref{fig:mf}. The MFs of the LI8020SA sensor at $10~\mathrm{keV}$ obtained from experiments at both SPring-8 and KEK-PF exhibit consistency, suggesting that the polarization degree of the SPring-8 beam was nearly $100\%$. The uncertainty in the MFs was assessed by considering the uncertainty in the polarization degree of the beam.

We also plot the MFs of the GMAX0505RF sensor in Fig. \ref{fig:mf} for comparison. The GMAX0505RF, manufactured by GPixel Inc., is a CMOS image sensor designed for visible light and near-infrared imaging. Its imaging area consists of $5120 \times 5120$ pixels, with each pixel measuring $2.5~{\rm \mu m} \times 2.5~{\rm \mu m}$. Since this sensor is designed to improve sensitivity to near-infrared light, we anticipate that the detection layer thickness of GMAX0505RF is lager than that of LI8020SA. The analysis procedure of the GMAX0505RF can be found in Odaka et al. (in prep). The MFs of the LI8020SA, with a smaller pixel size, exceeded those of the GMAX0505RF at all energies. Additionally, MFs increased with energy.

\begin{table}[htb]
\caption{The modulation factors of the LI8020SA sensor.}
\begin{center}
\begin{tabular}{ c c c }
\hline
\hline
 Energy       &    \multicolumn{2}{c}{ Modulation factor}              \\
              &       KEK-PF   &   SPring-8    \\
\hline
$6~{\rm keV}$ &       $3.77\% \pm 0.20\%$   &                      \\
$10~{\rm keV}$ &       $8.02\% \pm 0.25\%$   &   $8.32\% \pm 0.34\%$  \\
$12~{\rm keV}$ &                         &    $10.90\% \pm 0.43\%$  \\
$16~{\rm keV}$ &                         &    $13.66\% \pm 0.65\%$  \\
$22~{\rm keV}$ &                         &    $16.10\% \pm 0.68\%$  \\
\hline
\end{tabular}
\label{tb:mf}
\end{center}
\end{table}

\subsection{Quantum efficiency}
\label{sub_qe}

We compared the count rates of the CMOS sensor with those of the SDD to estimate the QE of the CMOS sensor. To determine the number of detected events by the CMOS sensor, we excluded events originating from detected fluorescence lines and higher harmonic beams, based on their reconstructed energy and multiplicity values. 
The criteria used for event selection can influence the count rates and thus introduce uncertainty. To obtain the number of events and their uncertainties, we selected smaller and larger regions in the (reconstructed energy, multiplicity) space at each energy. The number of events was computed by averaging the events within these regions. Dividing by the total exposure time yielded the count rates of the CMOS sensor. The uncertainty in the counts was estimated as half of the difference in counts between the regions.

The output count rate of the SDD data can be determined by dividing the number of events around the peak by the accumulation time. In the experiment, the input count rate was sufficiently high that the dead time could not be neglected. Assuming the paralyzable model, we reconstructed the input count rate by numerically solving the formula $R_\mathrm{out} = R_\mathrm{in} \exp\left(-R_\mathrm{in} T_\mathrm{dead} \right)$, where $T_\mathrm{dead} = 2(1+0.05)\left(T_\mathrm{peak} + T_\mathrm{flat} \right)$, with $T_\mathrm{peak}$ representing the peaking time and $T_\mathrm{flat}$ representing the flat top duration \cite{redus2008}. Taking into account the detection layer depth of the SDD ($500~\mathrm{\mu m}$) and the transmission ratio of the window (B4C), we estimated the input count rates of the CMOS sensor.

\begin{figure}[htb]
\begin{center}
\includegraphics[width=7cm]{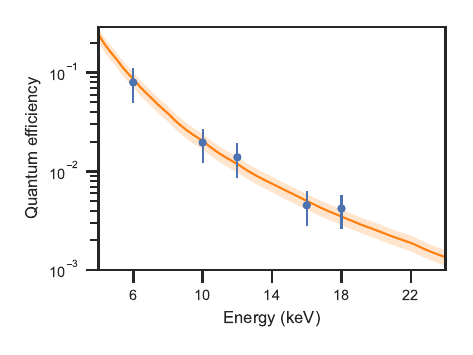}
\caption{Quantum efficiency of LI8020SA at each energy. The blue circles represent the measured quantum efficiency of the LI8020SA sensor. The orange line and shaded area represent the fitted silicon slab model with the best-fit parameter and the $1\sigma$ range for the detection layer thickness of $2.67 \pm 0.48~{\rm \mu m}$.}
\label{fig:qe_li8020}
\end{center}
\end{figure}

The QE of the CMOS sensor was determined based on the detected count rates and the estimated input count rates. The uncertainty in these quantum efficiencies was assessed using the variance of multiple pairs of data at $10~\mathrm{keV}$ and $12~\mathrm{keV}$. Assuming the sensor can approximated by a silicon slab, we derived the thickness of the detection layer from the QE of the CMOS sensor, as illustrated in Fig. \ref{fig:qe_li8020}. The inferred detection layer thickness was $2.67 \pm 0.48~{\rm \mu m}$. The detection layer thickness of the GMAX0505RF is $18.1 \pm 3.3~{\rm \mu m}$ (Odaka et al. in prep). As anticipated, the NIR-enhanced RF sensor exhibited a thicker detection layer.

\section{Discussion}
\label{sec_discussion}

\subsection{Polarization sensitivity}
\label{sub_qf}

\begin{figure}[htb]
\begin{center}
\includegraphics[width=7cm]{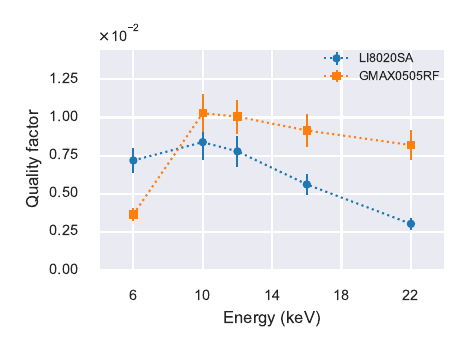}
\caption{Quality factors of the CMOS sensors at each energy. The quality factors of the GMAX0505RF sensor were obtained from Odata et al. (in prep).}
\label{fig:quality factor}
\end{center}
\end{figure}

The sensitivity of the polarimeter is often represented as $\mathrm{MDP}_{99}$, the minimum detectable polarization at the $99\%$ confidence level. $\mathrm{MDP}_{99}$ can be calculated using the equation:
\begin{equation}
\mathrm{MDP}_{99} \simeq \frac{4.29}{\mu\sqrt{N}}
\label{eq:MDP99}
\end{equation}
where $\mu$ represents the MF and $N$ represents the number of detected source events. For $N = \varepsilon A F T$, where $A$, $F$, and $T$ denote the area, source flux, and observation time, respectively, the $\mathrm{MDP}_{99}$ is determined by a quality factor, $q = \mu\sqrt{\epsilon}$, of the sensor. Figure \ref{fig:quality factor} illustrates the quality factors of the LI8020SA and GMAX0505RF sensors at various energies. In this calculation, we employed a multiplicity range of $2$--$81$ for LI8020SA data and $2$--$4$ for GMAX0505RF data at $6$, $10$, and $12~\mathrm{keV}$, $2$--$10$ at $16~\mathrm{keV}$, and $2$--$25$ at $22~\mathrm{keV}$. Above $10~\mathrm{keV}$, the quality factor of the GMAX0505 sensor exceeds that of the LI8020SA due to its higher efficiency.
Conversely, below approximately $10~\mathrm{keV}$, the quality factor of the LI8020SA surpasses that of the GMAX0505RF. Therefore, the LI8020SA can offer better polarization sensitivity for sources that are highly polarized and bright below $10~\mathrm{keV}$. Consequently, the LI8020SA provides better polarization sensitivity for sources that are highly polarized and bright below $10~\mathrm{keV}$.

The MF of the LI8020SA exceeds that of the GMAX0505RF in the energy range of $6$--$22~\mathrm{keV}$. This is attributed to the smaller pixel size of the LI8020SA and the fact that the impact of thermal diffusion does not significantly diminish the advantage of the smaller pixel size. This suggests that a smaller pixel size is crucial for improving the MF of the sensor. However, the quality factor of the LI8020SA was lower than that of the GMAX0505RF above $10~\mathrm{keV}$ due to the thinner detection layer thickness. Therefore, for the development of a more sensitive polarimeter, a sensor with a thicker detection layer is desirable. We propose the use of CMOS sensors for X-ray polarimetry in bright sources with significant exposure times. The feasibility of X-ray polarimetry using the GMAX0505RF will be discussed in Odaka et al. (in prep).

\subsection{Comparison with simulations}
\label{sub_cf_simulation}

The MF of detectors is influenced by sensor properties such as pixel size and thermal diffusion. To investigate whether the effects of pixel size and thermal diffusion can account for the measured MF of the sensor, we conducted simulations of a silicon pixel sensor. For this purpose, we utilized ComptonSoft, a simulation software based on Geant4. In our simulation setup, we approximated the detector as a slab of silicon with a thickness of $2.7~\mathrm{\mu m}$, consistent with the experimental result in subsection \ref{sub_qe}. 
We did not include the insensitive neutral region, likely present on the back side of the sensor, in the mass model used in the simulation. The detector was modeled to have $256\times256$ pixels, with each pixel measuring $1.5~\mathrm{\mu m}\times 1.5~\mathrm{\mu m}$. Perfectly polarized incident photons were emitted to irradiate a central $1.5~\mathrm{\mu m}\times 1.5~\mathrm{\mu m}$ rectangle region of the sensor. We simulated photon energies of $6$, $10$, $12$, $16$, and $22~\mathrm{keV}$ and polarization angles of $0^{\circ}$, $45^{\circ}$, and $90^{\circ}$, mimicking the conditions of the beam experiment. The deposited energies (charges) and their positions were simulated with Geant4. To simulate the diffusion effect, the positions of the charges were rearranged, following Gaussian distributions centered at their original positions with a spread of $\sigma$. The rearranged charges in each pixel were combined and assigned to the corresponding pixel. To introduce noise, the charge value for each pixel was randomized using a Gaussian function. The Gaussian width was based on the energy resolution estimated from events with a multiplicity of 1. The pixel energy values were converted to ADU values using the results of gain calibration. We extracted events and calculated the MF using the same method and thresholds as in the analysis of the beam experiment data. The events used for the analysis were within the multiplicity range of $2$--$81$.

\begin{figure}[htb]
\begin{center}
\includegraphics[width=7cm]{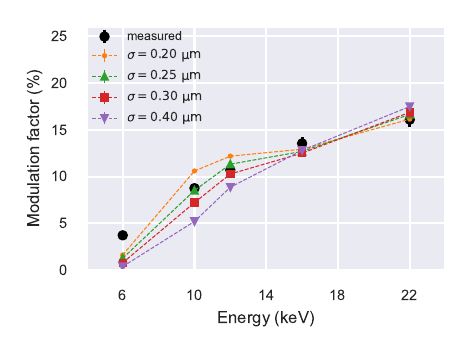}
\caption{Modulation factor of the simulation results at each energy for varying diffusion widths, $\sigma$. The sensor thick is fixed at $2.7~\mathrm{\mu m}$ with a pixel size of $1.5~\mathrm{\mu m}$.}
\label{fig:mf_sim_sigma}
\end{center}
\end{figure}

For simplicity, we applied the same $\sigma$ value to all hits, regardless of their initial depth. We then simulated several $\sigma$ values and calculated MFs. The sensor thickness was set to $2.7~\mathrm{\mu m}$, as measured in subsection \ref{sub_qe}, with a pixel size of $1.5~\mathrm{\mu m}$. The results are shown in Fig. \ref{fig:mf_sim_sigma}. We also plotted the measured MF obtained using the events within the multiplicity range of $2$--$81$. It can be observed that the MF below $\lesssim 12~\mathrm{keV}$ increases as $\sigma$ decreases. This is likely because polarization information in the photoelectron track tends to be better preserved under lower diffusion effects. We also verified that the measured MFs could be reproduced with $\sigma = 0.25~\mathrm{\mu m}$ except for the MF at $6~\mathrm{keV}$. This condition corresponds to the constant doping concentration of $\sim 5\times 10^{14}~\mathrm{cm}^{-3}$, a reasonable value.

\begin{figure}[htb]
\begin{center}
\includegraphics[width=7cm]{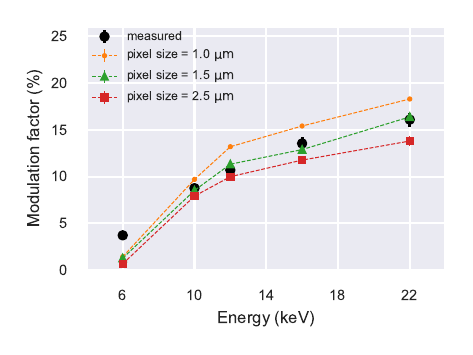}
\caption{Modulation factor of the simulation results at each energy for varying pixel size. The $\sigma$ is fixed at $0.25~\mathrm{\mu m}$ with a sensor thick of $2.7~\mathrm{\mu m}$.}
\label{fig:mf_sim_pixsize}
\end{center}
\end{figure}

\begin{figure}[htb]
\begin{center}
\includegraphics[width=7cm]{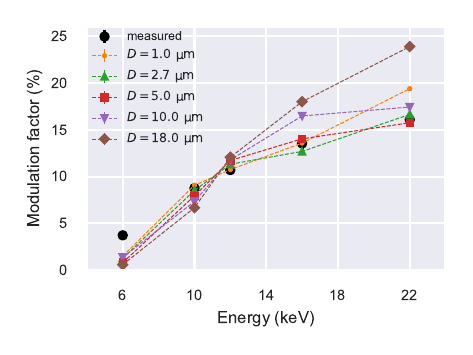}
\caption{Modulation factor of the simulation results at each energy for varying sensor thickness $D$. The $\sigma$ is fixed at $0.25~\mathrm{\mu m}$ with a pixel size of $1.5~\mathrm{\mu m}$.}
\label{fig:mf_sim_dl}
\end{center}
\end{figure}

To investigate the dependence of the MF on sensor parameters, we conducted simulations varying the sensor pixel size and thickness, maintaining $\sigma = 0.25~\mathrm{\mu m}$. When varying the pixel size, the thickness was fixed at $2.7~\mathrm{\mu m}$, and the X-ray irradiation area was adjusted to match the pixel size. We kept the number of pixels at $256\times256$. When varying the thickness, the pixel size was fixed at $1.5~\mathrm{\mu m}$. The simulation results are shown in Fig. \ref{fig:mf_sim_pixsize} and Fig. \ref{fig:mf_sim_dl}. Fig. \ref{fig:mf_sim_pixsize} illustrates an enhancement in the MF with a decrease in pixel size. This is likely because a smaller pixel sensor allows for more precise tracking of the photoelectron. 
Fig. \ref{fig:mf_sim_dl} shows the dependence of the MF on the sensor thickness $D$. The MF at energies above $\gtrsim 16~\mathrm{keV}$ increases with increasing $D$ if $D \gtrsim 5.0~\mathrm{\mu m}$, which is approximately equal to the range of $20~\mathrm{keV}$ photoelectron. In thinner sensors, photoelectrons are more likely to escape from the sensor, altering the ratio of events originating from photoelectrons exhibiting non-linear motion due to large-angle scattering.
This phenomenon could explain the MF's dependence on sensor thickness. Given that an increase in thickness also enhances QE, a substantial detector thickness emerges as a crucial factor in enhancing polarization sensitivity. The simulation results indicate that a more sensitive polarimeter could be developed by using a sensor with a thicker detection layer, smaller pixel size, and reduced thermal diffusion effect.

\section{Conclusion}
\label{sec_conclusion}

We are developing an imaging polarimeter by combining a fine-pixel CMOS image sensor with a coded aperture mask as part of the cipher project, aiming to achieve X-ray polarimetry in the energy range of $10$--$30~\mathrm{keV}$. We have successfully conducted a proof-of-concept experiment using a fine-pixel CMOS sensor with a pixel size of $2.5~\mathrm{\mu m}$. In this study, we performed beam experiments to evaluate the MF of the CMOS sensor with a pixel size of $1.5~\mathrm{\mu m}$ manufactured by Canon and investigated if the MF was improved. The measured MF of the smaller pixel sensor was $8.32\% \pm 0.34\%$ at $10~\mathrm{keV}$ and $16.10\% \pm 0.68\%$ at $22~\mathrm{keV}$, exceeding those of the $2.5~\mathrm{\mu m}$ sensor in the $6$--$22~\mathrm{keV}$ range. To develop a more sensitive polarimeter, a sensor with a thicker detection layer, smaller pixel size, and reduced thermal diffusion effect is desirable.

\section*{Acknowledgment}
We are grateful for the support of T. Tamagawa, T. Kohmura, and N. Narukage, and the technical assistance from H. Sagayama at KEK and M. Hoshino and K. Uesugi at SPring-8. 
This research was supported by FoPM, WINGS Program, the University of Tokyo (TI), Research Fellowships of Japan Society for the Promotion of Science (JSPS) for Young
Scientists (TI), JSPS KAKENHI Grant No. 23KJ0780 (TI), 21K13963 (KH), 18H05861, 19H01906, 19H05185, 22H00128, and 22K18277 (HO).
The synchrotron radiation experiments were performed at BL20B2 in SPring-8 with the approval of the Japan Synchrotron Radiation Research Institute (JASRI) (Proposal No. 2019B1369, 2020A1343, 2022B1477, and 2023A1476). This work was performed under the approval of the Photon Factory Program Advisory Committee (Proposal No. 2023G672). During the preparation of this work the authors used ChatGPT 3.5 in order to improve readability and language. After using this tool, the authors reviewed and edited the content as needed and take full responsibility for the content of the publication.

\bibliography{nima}

\begin{thebibliography}{10}
\expandafter\ifx\csname url\endcsname\relax
  \def\url#1{\texttt{#1}}\fi
\expandafter\ifx\csname urlprefix\endcsname\relax\def\urlprefix{URL }\fi
\expandafter\ifx\csname href\endcsname\relax
  \def\href#1#2{#2} \def\path#1{#1}\fi

\bibitem{chauvin2017}
M.~{Chauvin}, H.~G. {Flor{\'e}n}, M.~{Friis}, et~al., {Shedding new light on the Crab with polarized X-rays}, Scientific Reports 7 (2017) 7816.
\newblock \href {http://arxiv.org/abs/1706.09203} {\path{arXiv:1706.09203}}, \href {http://dx.doi.org/10.1038/s41598-017-07390-7} {\path{doi:10.1038/s41598-017-07390-7}}.

\bibitem{hitomi2018}
{Hitomi Collaboration}, F.~{Aharonian}, H.~{Akamatsu}, et~al., {Detection of polarized gamma-ray emission from the Crab nebula with the Hitomi Soft Gamma-ray Detector}, PASJ 70~(6) (2018) 113.
\newblock \href {http://arxiv.org/abs/1810.00704} {\path{arXiv:1810.00704}}, \href {http://dx.doi.org/10.1093/pasj/psy118} {\path{doi:10.1093/pasj/psy118}}.

\bibitem{weisskopf2022}
M.~C. {Weisskopf}, P.~{Soffitta}, L.~{Baldini}, et~al., {The Imaging X-Ray Polarimetry Explorer (IXPE): Pre-Launch}, Journal of Astronomical Telescopes, Instruments, and Systems 8~(2) (2022) 026002.
\newblock \href {http://arxiv.org/abs/2112.01269} {\path{arXiv:2112.01269}}, \href {http://dx.doi.org/10.1117/1.JATIS.8.2.026002} {\path{doi:10.1117/1.JATIS.8.2.026002}}.

\bibitem{beilicke2014}
M.~{Beilicke}, F.~{Kislat}, A.~{Zajczyk}, Q.~{Guo}, R.~{Endsley}, M.~{Stork}, R.~{Cowsik}, P.~{Dowkontt}, S.~{Barthelmy}, T.~{Hams}, T.~{Okajima}, M.~{Sasaki}, B.~{Zeiger}, G.~{de Geronimo}, M.~G. {Baring}, H.~{Krawczynski}, {Design and Performance of the X-ray Polarimeter X-Calibur}, Journal of Astronomical Instrumentation 3~(2) (2014) 1440008.
\newblock \href {http://arxiv.org/abs/1412.6457} {\path{arXiv:1412.6457}}, \href {http://dx.doi.org/10.1142/S225117171440008X} {\path{doi:10.1142/S225117171440008X}}.

\bibitem{abarr2020}
Q.~{Abarr}, M.~{Baring}, B.~{Beheshtipour}, et~al., {Observations of a GX 301-2 Apastron Flare with the X-Calibur Hard X-Ray Polarimeter Supported by NICER, the Swift XRT and BAT, and Fermi GBM}, \apj 891~(1) (2020) 70.
\newblock \href {http://arxiv.org/abs/2001.03581} {\path{arXiv:2001.03581}}, \href {http://dx.doi.org/10.3847/1538-4357/ab672c} {\path{doi:10.3847/1538-4357/ab672c}}.

\bibitem{odaka2020}
H.~{Odaka}, T.~{Kasuga}, K.~{Hatauchi}, et~al., {Concept of a CubeSat-based hard x-ray imaging polarimeter: cipher}, in: J.-W.~A. {den Herder}, S.~{Nikzad}, K.~{Nakazawa} (Eds.), Space Telescopes and Instrumentation 2020: Ultraviolet to Gamma Ray, Vol. 11444 of Society of Photo-Optical Instrumentation Engineers (SPIE) Conference Series, 2020, p. 114445V.
\newblock \href {http://dx.doi.org/10.1117/12.2560615} {\path{doi:10.1117/12.2560615}}.

\bibitem{kasuga2020}
T.~Kasuga, H.~Odaka, K.~Hatauchi, et~al., Artifact-less coded aperture imaging in the x-ray band with multiple different random patterns, Journal of Astronomical Telescopes, Instruments, and Systems 6~(3) (2020) 035002--035002.

\bibitem{goto2001}
S.~{Goto}, K.~{Takeshita}, Y.~{Suzuki}, et~al., {Construction and commissioning of a 215-m-long beamline at SPring-8}, Nuclear Instruments and Methods in Physics Research A 467~(2001) (2001) 682--685.
\newblock \href {http://dx.doi.org/10.1016/S0168-9002(01)00445-4} {\path{doi:10.1016/S0168-9002(01)00445-4}}.

\bibitem{asakura2019}
K.~{Asakura}, K.~{Hayashida}, T.~{Hanasaka}, et~al., {X-ray imaging polarimetry with a 2.5-{\ensuremath{\mu}}m pixel CMOS sensor for visible light at room temperature}, Journal of Astronomical Telescopes, Instruments, and Systems 5 (2019) 035002.
\newblock \href {http://arxiv.org/abs/1906.00012} {\path{arXiv:1906.00012}}, \href {http://dx.doi.org/10.1117/1.JATIS.5.3.035002} {\path{doi:10.1117/1.JATIS.5.3.035002}}.

\bibitem{agostinelli2003}
S.~{Agostinelli}, J.~{Allison}, K.~{Amako}, et~al., {GEANT4{\textemdash}a simulation toolkit}, Nuclear Instruments and Methods in Physics Research A 506~(3) (2003) 250--303.
\newblock \href {http://dx.doi.org/10.1016/S0168-9002(03)01368-8} {\path{doi:10.1016/S0168-9002(03)01368-8}}.

\bibitem{allison2006}
J.~{Allison}, K.~{Amako}, J.~{Apostolakis}, et~al., {Geant4 developments and applications}, IEEE Transactions on Nuclear Science 53~(1) (2006) 270--278.
\newblock \href {http://dx.doi.org/10.1109/TNS.2006.869826} {\path{doi:10.1109/TNS.2006.869826}}.

\bibitem{allison2016}
J.~{Allison}, K.~{Amako}, J.~{Apostolakis}, et~al., {Recent developments in GEANT4}, Nuclear Instruments and Methods in Physics Research A 835 (2016) 186--225.
\newblock \href {http://dx.doi.org/10.1016/j.nima.2016.06.125} {\path{doi:10.1016/j.nima.2016.06.125}}.

\bibitem{bellazzini2003}
R.~{Bellazzini}, F.~{Angelini}, L.~{Baldini}, et~al., {Novel gaseus X-ray polarimeter: data analysis and simulation}, in: S.~{Fineschi} (Ed.), Polarimetry in Astronomy, Vol. 4843 of Society of Photo-Optical Instrumentation Engineers (SPIE) Conference Series, 2003, pp. 383--393.
\newblock \href {http://dx.doi.org/10.1117/12.459381} {\path{doi:10.1117/12.459381}}.

\bibitem{redus2008}
R.~H. Redus, A.~C. Huber, D.~J. Sperry, Dead time correction in the dp5 digital pulse processor, in: 2008 IEEE Nuclear Science Symposium Conference Record, 2008, pp. 3416--3420.
\newblock \href {http://dx.doi.org/10.1109/NSSMIC.2008.4775075} {\path{doi:10.1109/NSSMIC.2008.4775075}}.

\end{thebibliography}
\end{document}